# Low-loss composite photonic platform based on 2D semiconductor monolayers

Ipshita Datta[1,*], Sang Hoon Chae[2,*], Gaurang R. Bhatt[1], Mohammad A. Tadayon[1], Baichang Li[2], Yiling Yu[3], Chibeom Park[4], Jiwoong Park[4], Linyou Cao[3], D. N. Basov[5], James Hone[2] and Michal Lipson[1]

[1]Department of Electrical Engineering, Columbia University, New York, New York 10027, USA

[2]Department of Mechanical Engineering, Columbia University, New York, New York 10027, USA

[3]Department of Materials Science and Engineering and Department of Physics, North Carolina State University, Raleigh, North Carolina 27695, USA

[4]Department of Chemistry, Institute for Molecular Engineering, James Franck Institute, University of Chicago, Chicago, IL 60637, USA

[5]Department of Physics, Columbia University, New York, New York 10027, USA

Corresponding Author – ml3745@columbia.edu

*These authors contributed equally to this work.

Two dimensional (2D) materials such as graphene and transition metal dichalcogenides (TMDs) are promising for optical modulation, detection, and light emission since their material properties can be tuned on-demand via electrostatic doping[1–18]. The optical properties of TMDs have been shown to change drastically with doping in the wavelength range near the excitonic resonances[19–22]. However, little is known about the effect of doping on the optical properties of TMDs away from these resonances, where the material is transparent and therefore could be leveraged in photonic circuits. Here, we probe the electro-optic response of monolayer TMDs at near infrared (NIR) wavelengths (i.e. deep in the transparency regime), by integrating them on silicon nitride (SiN) photonic structures to induce strong light-matter interaction with the monolayer. We dope the monolayer to carrier densities of $(7.2 \pm 0.8) \times 10^{13}$ cm$^{-2}$, by electrically gating the TMD using an ionic liquid [P14$^+$] [FAP$^-$]. We show strong electro-refractive response in monolayer tungsten disulphide (WS$_2$) at NIR wavelengths by measuring a large change in the real part of refractive index $\Delta n = 0.53$, with only a minimal change in the imaginary part $\Delta k = 0.004$. The doping induced phase change ($\Delta n$), compared to the induced absorption ($\Delta k$) measured for WS$_2$ ($\Delta n/\Delta k \sim 125$), a key metric for photonics, is an order of magnitude higher than the $\Delta n/\Delta k$ for bulk materials like silicon ($\Delta n/\Delta k \sim 10$)[23], making it ideal for various photonic applications[24–28]. We further utilize this strong tunable effect to demonstrate an electrostatically gated SiN-WS$_2$ phase modulator using a WS$_2$-HfO$_2$ (Hafnia)-ITO (Indium Tin Oxide) capacitive configuration, that achieves a phase modulation efficiency ($V_\pi L$) of 0.8 V · cm with a RC limited bandwidth of 0.3 GHz.

In order to probe the doping induced electro-optic response of TMDs, we utilize a SiN-TMD composite waveguide platform in which the optical mode of the composite waveguide is shared between the TMD monolayer and the dielectric SiN waveguide. These waveguides are then incorporated into photonic structures such as microring resonators which are sensitive to small changes in absorption and phase induced by electrostatic doping of the TMD monolayer. The $SiO_2$-clad SiN waveguides[10] are fabricated using standard techniques, and then planarized to permit mechanical transfer of TMD layers and subsequent lithographic patterning to define Ti/Au contacts for gating (see Methods). We probe the response of these structures in the NIR region ($\lambda$ ~ 1550 nm), far below the excitonic resonances of the TMDs (~ 620 nm for $WS_2$; 660 nm for $MoS_2$).

We first characterize the electro-optic response of monolayer $WS_2$ by measuring the effect of gating on the response of a microring resonator cavity (1.3 µm × 330 nm passive SiN waveguide clad with 100 nm of $SiO_2$, embedded in a ring of 60 µm radius), as depicted in figure 1a. The high-Q (~120,000) cavity is critically coupled to a SiN bus waveguide, such that the transmission spectrum is extremely sensitive to small changes in phase and absorption within the cavity. A 30 µm arc length of $WS_2$ grown by metal-organic chemical vapor deposition[29] (MOCVD) is patterned onto the ring and electrically contacted (see Methods). We tune the effective index of the optical mode by electrostatically doping the monolayer using an ionic liquid [P14$^+$] [FAP$^-$], chosen due to its stability to gating under atmospheric exposure at room temperature[30]. As shown in figure 1b, we apply a potential difference between the electrode connected to $WS_2$ and a nearby counter-electrode, which results in the accumulation of ions at the surface of $WS_2$. We note that ionic liquid gated devices are dominated by the quantum capacitance ($C_q$)[30], such that no carriers are injected to the $WS_2$ within the bandgap region (roughly -1 to 1 V). The maximum electron doping density induced in the monolayer is about 7.2 ± 0.8 x $10^{13}$ cm$^{-2}$ with an applied voltage of 2 V (see Supplementary Section I). One can see from the cross-section of the platform in figure 1c, that the optical mode is shared between the monolayer and SiN waveguide. We estimate an optical mode overlap of 0.06 % using COMSOL Multiphysics simulations (see Methods).

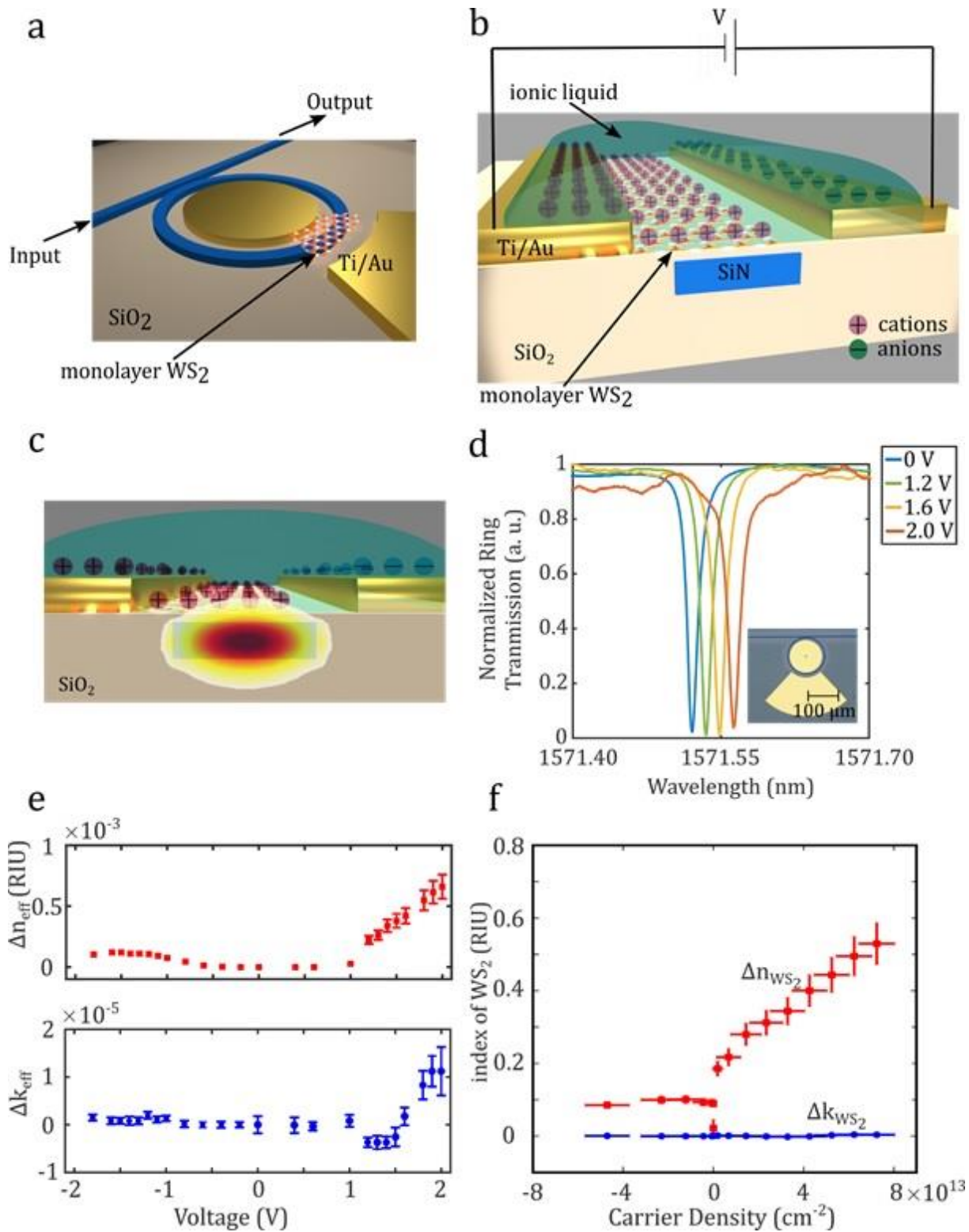

**Figure 1 | Schematic of the ionic liquid gated SiN-WS₂ platform.** (a) Schematic representation of a microring resonator with patterned monolayer WS$_2$ and Ti/Au electrodes, in which only one of the electrodes is in contact with the monolayer WS$_2$. (b) Illustration of the composite SiN-WS$_2$ waveguide with ionic liquid ([P14$^+$] [FAP$^-$]) cladding.

The monolayer WS$_2$ is doped by applying a bias voltage across the two electrodes through the ionic liquid, resulting in the accumulation of charged carriers at the interface of WS$_2$, thereby creating image charges in the monolayer. (c) TE mode profile of the propagating mode in the SiN waveguide with 0.06 % overlap with monolayer WS$_2$. (d) Normalized transmission response of the microring resonator for different voltages applied across the two electrodes. One can see that the resonance of the microring red shifts with significantly smaller broadening of the resonance with increasing voltage, indicating that the propagating mode undergoes strong phase shift but with minimal absorption. Bottom right inset is the optical micrograph of the fabricated ring modulator with SiN-WS$_2$ waveguides, before ionic liquid is dispersed on the devices. (e) Change in real and imaginary part of the effective index ($\Delta n_{\text{eff}}$ and $\Delta k_{\text{eff}}$) of the composite SiN-WS$_2$ waveguide at different voltages, extracted from the normalized transmission spectra in figure 1d. The error bars shown in figure 1e account for the root mean square (rms) error from the numerical fit and a ± 5 μm error in the patterned length of the monolayer. (f) Change in the real and imaginary part of the refractive index of monolayer WS$_2$ ($\Delta n_{\text{WS2}}$ and $\Delta k_{\text{WS2}}$) with induced carrier densities, extracted from the change in effective index in figure 1e. The vertical error bars in figure 1f incorporates the rms error in the effective index measurements from figure 1e, variation in the thickness of monolayer (± 0.05 nm), variation in the height of the cladding oxide by about ± 25 nm and about ± 0.02 (RIU) variation in the index of the ionic liquid included during the COMSOL simulations. The horizontal error bars accounts for the error in the intrinsic doping of the monolayer WS$_2$.

Figure 1d shows the measured transmission spectrum of the ring resonator as a function of voltage. The cavity has a resonance near 1571 nm, with narrow linewidth confirming that incorporation of WS$_2$ introduces negligible loss and does not degrade the quality factor of the microring resonator. At positive gate voltages above 1 V, we observe a red shift in the resonance wavelength, indicating an increase in the effective index of the resonator. Remarkably, the resonance linewidth is largely unchanged, thereby showing that doping does not introduce substantial loss. The measured changes in resonance wavelength and quality factor with applied voltage can be used to derive the changes in real and imaginary components of the effective index ($\Delta n_{\text{eff}}$ and $\Delta k_{\text{eff}}$, respectively) of the composite waveguide (see Supplementary Section II). Figure 1e shows the measured $\Delta n_{\text{eff}}$ and $\Delta k_{\text{eff}}$ as a function of gate voltage. $\Delta n_{\text{eff}}$ increases linearly with gate voltage with the onset of n-type doping (above 1V). In this same range, $\Delta k_{\text{eff}}$ is almost two orders of magnitude smaller. We can then determine the underlying change in real and imaginary part of the refractive index of WS$_2$, $\Delta n_{\text{WS2}}$ and $\Delta k_{\text{WS2}}$ respectively by COMSOL modelling of the monolayer as a 2D sheet with optical conductivity $\sigma_S$ integrated on a SiN waveguide (see Methods). Figure 1f shows $\Delta n_{\text{WS2}}$ and $\Delta k_{\text{WS2}}$ as a function of gate-induced carrier density in the monolayer WS$_2$. $\Delta n_{\text{WS2}}$ reaches 0.53 ± 0.06 RIU (refractive index units) for maximum doping of (7.2 ± 0.8) × 10$^{13}$ cm$^{-2}$, while $\Delta k_{\text{WS2}}$ is 0.004 ± 0.002. This indicates that monolayer WS$_2$ has a unique combination of strong electro-refractive response ($\Delta n_{WS2}$) and small electro-absorptive response ($\Delta k_{WS2}$) at NIR wavelengths, i.e. the propagating light undergoes significant phase change with low optical loss.

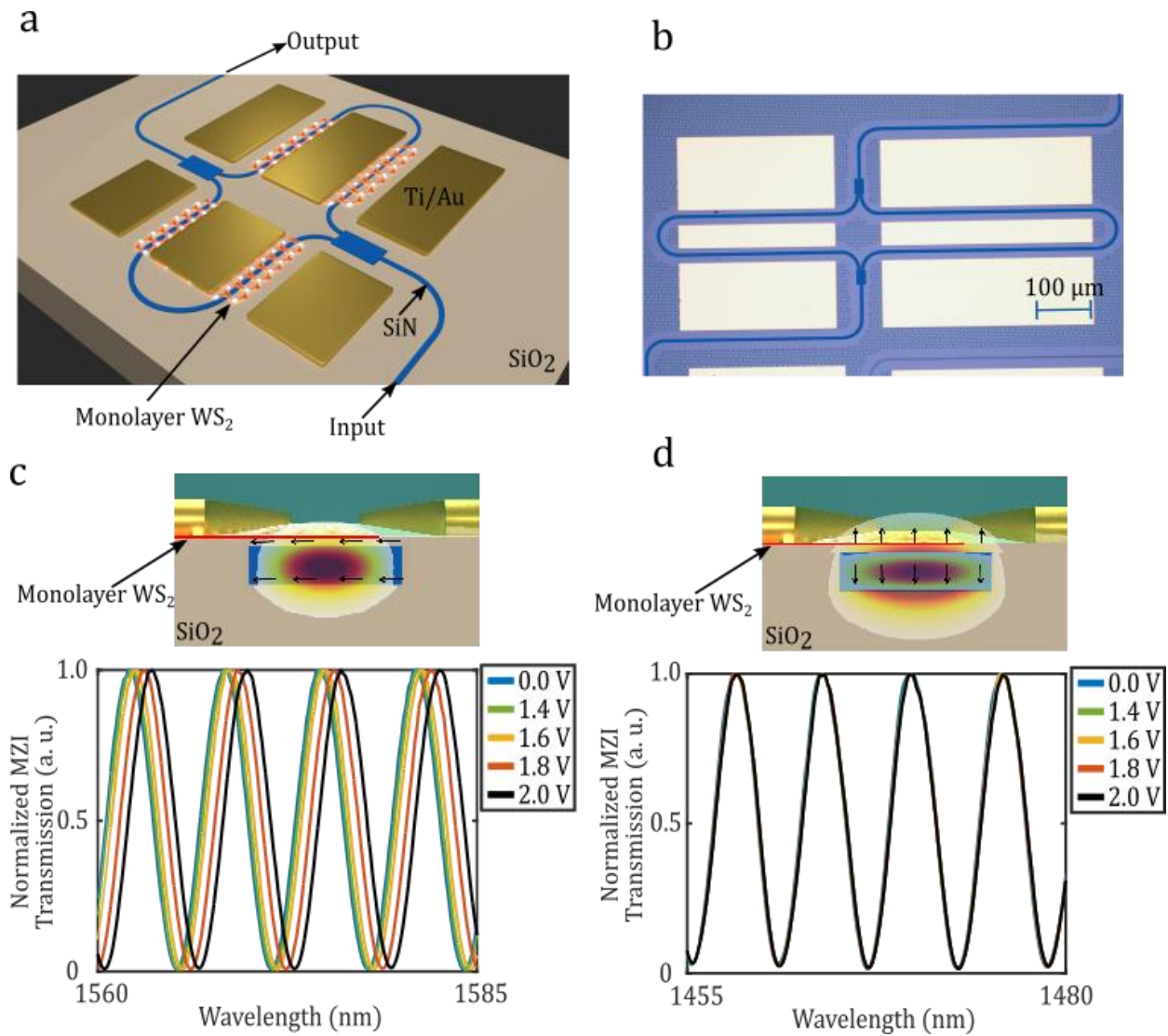

**Figure 2 | Tuning the effective index ($\Delta n_{eff}$) of the propagating TE and TM mode using monolayer WS₂.** (a) Diagrammatic illustration of an on-chip Mach Zehnder interferometer (MZI) with patterned monolayer WS₂ on both the arms of MZI. (b) Optical micrograph of the fabricated MZI with SiN-WS₂ waveguides, before ionic liquid is dispersed on the devices. (c) Normalized Transmission response of the MZI for the optical TE mode at different voltages applied across the two electrodes located on the longer arm of the MZI. One can see that the MZI transmission spectra exhibit fringes due to the path length difference between the two arms of the MZI, which red shifts with applied voltage. The top figure shows the optical mode profile for the TE mode, with the arrows illustrating the field lines aligned with the plane of monolayer WS₂. (d) Normalized Transmission response of the MZI for the optical TM mode

at different voltages. The bottom right inset shows the optical mode profile for the TM mode with the optical field lines perpendicular to the plane of monolayer $WS_2$.

We show that the strong phase change occurs only when the polarization of the propagating light is in plane with the monolayer TMD (i.e. TE mode). In order to probe the polarization effect, we embed the SiN-$WS_2$ composite waveguide in the arms of a Mach Zehnder interferometer (MZI), optimized to operate both for the TM and TE mode in the S wavelength band (1460 – 1490 nm) and C/L band (1530 - 1630 nm), respectively (see Methods). Figure 2a illustrates the diagrammatic representation of the MZI device while figure 2b shows an optical micrograph of the fabricated device. The MZI is designed with a length imbalance of 200 μm between the two arms such that the transmission spectra exhibit fringes with a visibility of 0.98 in the wavelength range spanning from 1560 - 1620 nm for TE mode and 1455 - 1480 nm for TM mode. A 500 μm long film of monolayer $WS_2$ is patterned on both the arms of the MZI, followed by the deposition of metal electrodes. We dope the monolayer by applying a voltage between the metal electrodes through the ionic liquid [$P14^+$] [$FAP^-$]. We estimate from COMSOL Multiphysics simulation that the optical mode overlap with the monolayer is 0.06% and 0.04% for the TE and TM mode, respectively (see Methods). In Figure 2c and 2d, we show the interference pattern at the MZI output for the TE and TM mode, respectively with different voltages applied to dope the monolayer $WS_2$ on the longer arm of the MZI. There is a pronounced wavelength shift in the MZI spectra for the TE mode (Figure 2c), as compared to minimal wavelength shift for the TM mode (Figure 2d), indicating that the doping-induced electro-refractive effect is strong only when the optical E field is aligned with the monolayer, despite similar mode overlap for the TE/TM mode with the monolayer. We further confirm that the doping of ionic liquid in the absence of the monolayer does not influence the MZI transmission response (see Supplementary Section III).

To leverage the doping-induced strong electro-refractive response in monolayer TMDs for photonic applications, we develop a fully integrated SiN-TMD platform that gates the monolayer TMD using a parallel plate capacitor configuration. We replace the ionic liquid used above with a stack of $HfO_2$ (hafnia) and transparent conducting oxide ITO (indium tin oxide) to form the TMD-$HfO_2$-ITO capacitor on the SiN waveguide, as illustrated in figure 3a. Standard processes are used to define the optimized 1 μm × 360 nm LPCVD SiN waveguide, clad with 260 nm of $SiO_2$ and a

500 µm long $WS_2$-$HfO_2$-ITO capacitor fabricated on each arm of the MZI (see Methods). We replace the MOCVD-grown[29] monolayer $WS_2$ used in the ionic liquid experiments with CVD-grown[21] films for the capacitive devices (see Supplementary Section IV). Supplementary figure S4a shows the false-coloured optical micrograph of the fabricated SiN-TMD MZI. In this configuration quantum capacitance effects are negligible,[30] and we estimate a linear induced charge density of $0.37 \pm 0.05 \times 10^{12}$ cm$^{-2}$ per volt, based on the thickness (26 nm) and dielectric permittivity of the $HfO_2$. In contrast to the ionic liquid devices where the voltage swing is limited between -2 V to 2 V, the voltage swing in these capacitive devices is determined by the thickness of the dielectric ($HfO_2$) and is in range of {-8V, 9V}.

Figure 3b shows the interference pattern at the MZI output for different voltages applied across the $WS_2$-$HfO_2$-ITO capacitor on the longer arm of the MZI. As in the ionic liquid gated devices, we induce a strong change in phase with applied voltage, thereby designing a phase modulator with a modulation efficiency ($V_\pi L$) of 1.33 V · cm, coupled with minimal change in extinction. Figure 3c shows the gating induced $\Delta n_{eff}$ of the composite waveguide, extracted by measuring the wavelength shift of the MZI spectra (see Supplementary Section IV), which reaches $7\times10^{-4}$ RIU for a swing voltage of 8V. From the extremum point at -4 V in Figure 3c, we infer that the charge neutrality point for the monolayer $WS_2$ layer is at -4 V, which corresponds to an initial electron doping of $1.5 \pm 0.2 \times 10^{12}$ cm$^{-2}$. This initial doping is likely due to sulfur vacancies arising from CVD growth but can also arise from substrate effects or adsorbates introduced in the transfer process[31].

We further enhance the performance of the capacitive SiN-$WS_2$ platform by increasing the optical mode overlap with monolayer $WS_2$ using SU-8 photoresist as a high index cladding (see Supplementary figure S4b). In figure 3c, we show an increase in the gating induced $\Delta n_{eff}$ from $7.1 \times 10^{-4}$ RIU to $1.35 \times 10^{-3}$ RIU which reduces the $V_\pi L$ from 1.33 V · cm to 0.8 V · cm (see Supplementary Section V). We estimate that the mode overlap for the unclad device is 0.016 % and for the SU-8 clad device is 0.03 %, based on our simulations (see Methods). The $\Delta n_{eff}$ for the SU-8 clad device is almost double the $\Delta n_{eff}$ extracted for an unclad device, in agreement with the improvement in the mode overlap from our COMSOL computations. Considering the relatively small mode overlap with the monolayer CVD $WS_2$ in this present work (0.03 %), one could envision an even higher phase modulation efficiency using alternative waveguide geometries with different degrees of mode confinement such as slot waveguides and photonic crystals. We measure a 3 dB bandwidth of 0.3 GHz in our devices with minimal DC electrical power dissipation of 0.64 nW in the $WS_2$-$HfO_2$-ITO capacitor (see Supplementary Section VI and VII). The frequency response of the phase modulator is measured using a 40 GHz fast photodiode and an electrical vector network analyzer (VNA) at 1550 nm. The bandwidth is currently limited by the resistance of monolayer $WS_2$, and can be increased with improved electrical contacts, optimized geometry, and control over the initial doping density.

Finally, we demonstrate that the gating induced electro-refractive effect extends to other semiconductor TMDs by fabricating and characterizing a SiN-$MoS_2$ composite waveguide embedded in a MZI structure (see figure 4a). The measured $\Delta n_{\text{eff}}$ as a function of gate voltage is shown in Figure 4b, and reaches a maximum of $6.4 \times 10^{-4}$ (RIU) at 8 V (see Supplementary Section VIII). The phase modulation efficiency $V_\pi L$ is 1.7 V • cm. Figure 4c compares the index change ($\Delta n$) for $MoS_2$ and $WS_2$ with gate voltage. In both cases the index changes strongly, with the change in $WS_2$ roughly double that $MoS_2$.

The demonstrated strong light matter interaction in monolayer TMDs could open up doors for a range of novel applications with these 2D materials and enable highly reconfigurable photonic circuits with low optical loss and power dissipation. The 2D configuration in the composite platform lends itself to extremely high doping of the material. The induced phase change, relative to the induced absorption (i.e $\Delta n/\Delta k$ – a key metric for photonics) is about 125 at carrier doping densities of $7 \times 10^{13}$ cm$^{-2}$, about two orders of magnitude higher than in traditional bulk materials such as silicon. The low absorption in TMDs at such high doping densities is due to the limited phase space in a lower dimensional system. Traditional phase modulators based on either thermo-optic or induced electro-optic $\chi^{(2)}$ effects have similar low optical losses as our SiN-TMD device, but suffer from either high electrical power consumption and low operation bandwidth[32–35] or require a large device footprint and requires complex fabrication techniques[36–38], respectively. For large scale photonic systems, wafer scale integration of TMD materials with silicon photonics can be done either as a direct TMD growth process on silicon wafers[29,39,40], or as a post processing step where large wafer scale TMD films[41] are transferred on a silicon photonics platform, fabricated in a standard foundry. The low electrical power dissipation and moderate operation bandwidth in our SiN-TMD platform allows for the large scale integration of these phase modulators in numerous applications such as LIDAR, phased arrays, optical switching, quantum and optical neural networks.

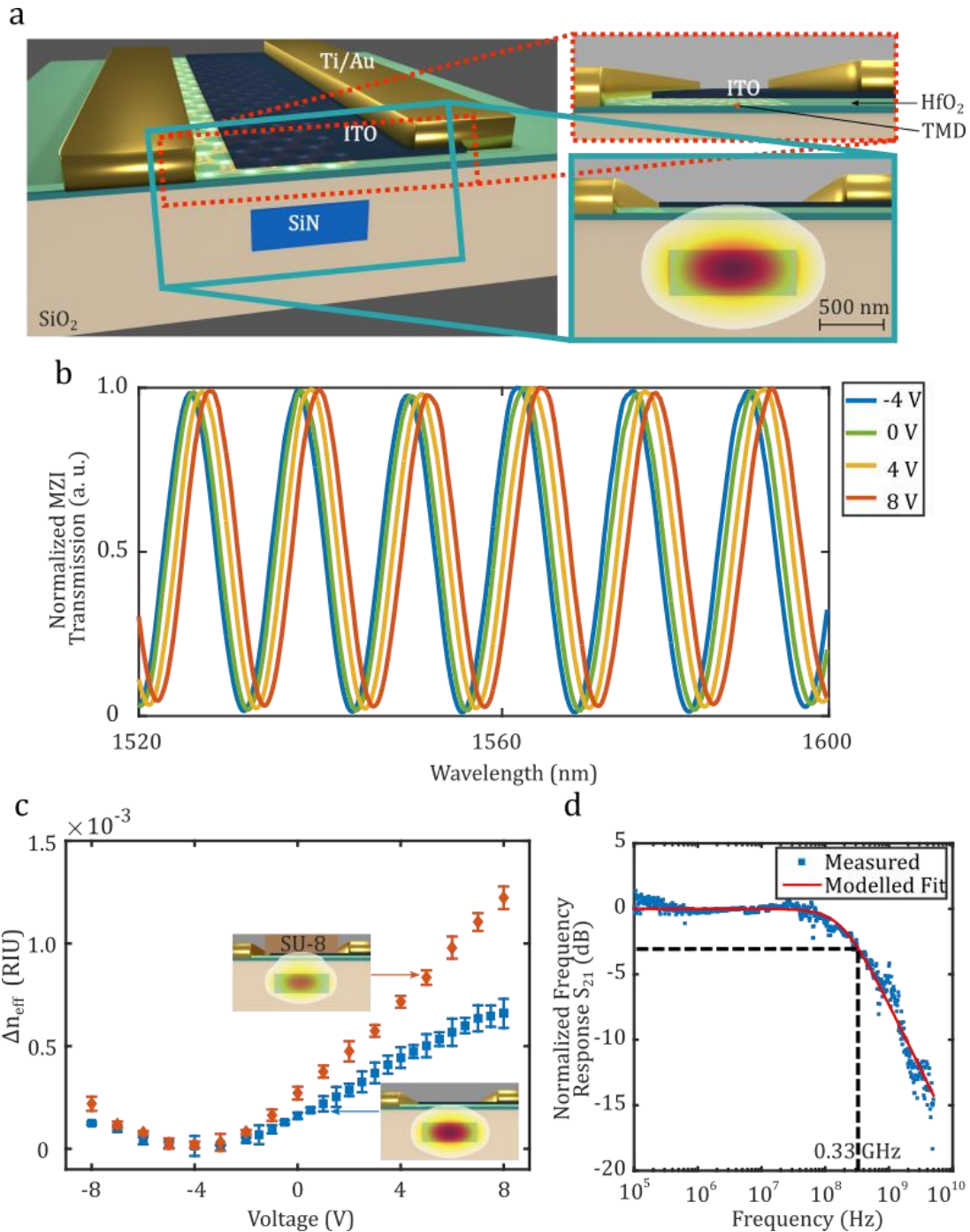

**Figure 3| Phase tuning of monolayer TMD using TMD-HfO$_2$-ITO capacitor.** (a) Illustration of a composite SiN-monolayer TMD waveguide, where the optical mode is shared between the SiN waveguide and TMD (bottom right).

The top right inset details the TMD-HfO$_2$-ITO parallel plate capacitor configuration which electrostatically dopes the monolayer WS$_2$, by applying a bias voltage between the two electrodes (TMD and ITO) across the dielectric HfO$_2$. (b) Normalized Transmission response of the MZI for different voltages applied across the WS$_2$-HfO$_2$-ITO capacitor. (c) Extracted change in effective index ($\Delta n_{\text{eff}}$) measured from the MZI spectra of the composite SiN-WS$_2$ waveguide with and without SU-8 cladding. The unclad device in figure 3a has a mode overlap of 0.016 % with the monolayer WS$_2$, while the SU-8 clad device has a mode overlap of 0.03 %. One can see that the enhanced mode overlap increases the maximum $\Delta n_{\text{eff}}$ of the composite SiN-WS$_2$ waveguide from $7 \times 10^{-4}$ to $1.3 \times 10^{-3}$ (corresponding to a $V_\pi L$ of 0.8 V·cm). The error bars show the variation in gating induced $\Delta n_{\text{eff}}$, extracted from multiple MZI devices with different lengths of WS$_2$-HfO$_2$-ITO capacitor. (d) Frequency response (S$_{21}$) of the unclad SiN-WS$_2$ phase modulator measured at 1550 nm.

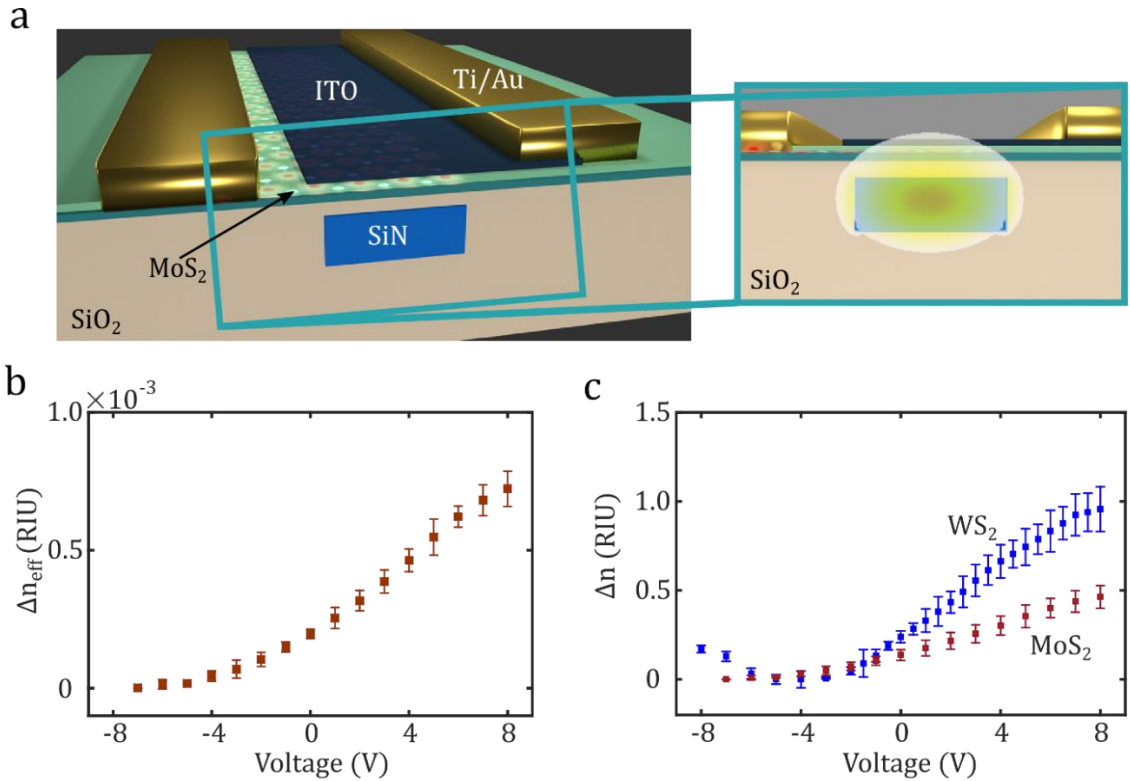

**Figure 4 | Phase tuning of monolayer MoS$_2$ in a composite SiN-MoS$_2$ waveguide using TMD-HfO$_2$-ITO capacitor.** (a) Illustration of a composite SiN-MoS$_2$ waveguide, where the optical mode is shared between the waveguide and MoS$_2$. Figure right inset illustrates the optical mode profile for the composite SiN-MoS$_2$ waveguide, with an optical mode overlap of 0.054% with the monolayer MoS$_2$. (b) Change in the effective index ($\Delta n_{\text{eff}}$) of the propagating mode with voltage, extracted from the normalized MZI transmission spectra of the composite SiN-MoS$_2$ waveguide (see Supplementary Section VIII). (c) Extracted change in refractive index of the monolayer TMD's ($\Delta n_{\text{TMD}}$) such as WS$_2$ and MoS$_2$ with a voltage swing of 8 V across the TMD-HfO$_2$-ITO capacitor. One can see that $\Delta n_{\text{WS2}} > \Delta n_{\text{MoS2}}$ even though the $\Delta n_{\text{eff}}$ for the unclad SiN-WS$_2$ composite waveguide (Figure 3c) is similar to $\Delta n_{\text{eff}}$ of the unclad SiN-MoS2 waveguide (Figure 4b), due to the higher mode overlap with the monolayer in the SiN-MoS$_2$ waveguide (0.054%) compared to that of the SiN-WS$_2$ waveguide (0.016%). The error bars in the $\Delta n$ calculations

include the rms error in the effective index measurements from Figure 3c and 4b, variation in the thickness of monolayer (± 0.05 nm), and variation in the height of the cladding oxide by about ± 25 nm included during the COMSOL simulations

## References

1. Bonaccorso, F., Sun, Z., Hasan, T. & Ferrari, A. C. Graphene photonics and optoelectronics. *Nat. Photonics* **4**, 611–622 (2010).
2. Wang, Q. H., Kalantar-Zadeh, K., Kis, A., Coleman, J. N. & Strano, M. S. Electronics and optoelectronics of two-dimensional transition metal dichalcogenides. *Nat. Nanotechnol.* **7**, 699–712 (2012).
3. Santos, E. J. G. & Kaxiras, E. Electrically Driven Tuning of the Dielectric Constant in $MoS_2$ Layers. *ACS Nano* **7**, 10741–10746 (2013).
4. Xia, F., Wang, H., Xiao, D., Dubey, M. & Ramasubramaniam, A. Two-dimensional material nanophotonics. *Nat. Photonics* **8**, 899–907 (2014).
5. Koppens, F. H. L. *et al.* Photodetectors based on graphene, other two-dimensional materials and hybrid systems. *Nat. Nanotechnol.* **9**, 780–793 (2014).
6. Ross, J. S. *et al.* Electrically tunable excitonic light-emitting diodes based on monolayer $WSe_2$ p–n junctions. *Nat. Nanotechnol.* **9**, 268–272 (2014).
7. Jariwala, D., Sangwan, V. K., Lauhon, L. J., Marks, T. J. & Hersam, M. C. Emerging Device Applications for Semiconducting Two-Dimensional Transition Metal Dichalcogenides. *ACS Nano* **8**, 1102–1120 (2014).
8. Schwarz, S. *et al.* Two-Dimensional Metal–Chalcogenide Films in Tunable Optical Microcavities. *Nano Lett.* **14**, 7003–7008 (2014).
9. C. Ferrari, A. *et al.* Science and technology roadmap for graphene, related two-dimensional crystals, and hybrid systems. *Nanoscale* **7**, 4598–4810 (2015).
10. Phare, C. T., Lee, Y.-H. D., Cardenas, J. & Lipson, M. Graphene electro-optic modulator with 30 GHz bandwidth. *Nat. Photonics* **9**, 511–514 (2015).
11. Pospischil, A. & Mueller, T. Optoelectronic Devices Based on Atomically Thin Transition Metal Dichalcogenides. *Appl. Sci.* **6**, 78 (2016).
12. Mak, K. F. & Shan, J. Photonics and optoelectronics of 2D semiconductor transition metal dichalcogenides. *Nat. Photonics* **10**, 216–226 (2016).
13. Xiao, J., Zhao, M., Wang, Y. & Zhang, X. Excitons in atomically thin 2D semiconductors and their applications. *Nanophotonics* **6**, 1309–1328 (2017).
14. Manzeli, S., Ovchinnikov, D., Pasquier, D., Yazyev, O. V. & Kis, A. 2D transition metal dichalcogenides. *Nat. Rev. Mater.* **2**, 17033 (2017).
15. Bernardi, M., Ataca, C., Palummo, M. & Grossman, J. C. Optical and Electronic Properties of Two-Dimensional Layered Materials. *Nanophotonics* **6**, 479–493 (2016).
16. Bie, Y.-Q. *et al.* A $MoTe_2$-based light-emitting diode and photodetector for silicon photonic integrated circuits. *Nat. Nanotechnol.* **12**, 1124–1129 (2017).
17. Basov, D. N., Averitt, R. D. & Hsieh, D. Towards properties on demand in quantum materials. *Nat. Mater.* **16**, 1077–1088 (2017).
18. Romagnoli, M. *et al.* Graphene-based integrated photonics for next-generation datacom and telecom. *Nat. Rev. Mater.* **3**, 392 (2018).

19. Ross, J. S. *et al.* Electrical control of neutral and charged excitons in a monolayer semiconductor. *Nat. Commun.* **4**, 1474 (2013).
20. Chernikov, A. *et al.* Electrical Tuning of Exciton Binding Energies in Monolayer $WS_2$. *Phys. Rev. Lett.* **115**, 126802 (2015).
21. Yu, Y. *et al.* Giant Gating Tunability of Optical Refractive Index in Transition Metal Dichalcogenide Monolayers. *Nano Lett.* **17**, 3613–3618 (2017).
22. Wang, G. *et al.* Colloquium: Excitons in atomically thin transition metal dichalcogenides. *Rev. Mod. Phys.* **90**, 21001 (2018).
23. Soref, R. & Bennett, B. Electrooptical effects in silicon. *IEEE J. Quantum Electron.* **23**, 123–129 (1987).
24. Doylend, J. K. *et al.* Two-dimensional free-space beam steering with an optical phased array on silicon-on-insulator. *Opt. Express* **19**, 21595–21604 (2011).
25. Sun, J., Timurdogan, E., Yaacobi, A., Hosseini, E. S. & Watts, M. R. Large-scale nanophotonic phased array. *Nature* **493**, 195–199 (2013).
26. Harris, N. C. *et al.* Quantum transport simulations in a programmable nanophotonic processor. *Nat. Photonics* **11**, 447–452 (2017).
27. Shen, Y. *et al.* Deep learning with coherent nanophotonic circuits. *Nat. Photonics* **11**, 441–446 (2017).
28. Miller, S. A. *et al.* 512-Element Actively Steered Silicon Phased Array for Low-Power LIDAR. in *Conference on Lasers and Electro-Optics (2018), paper JTh5C.2* JTh5C.2 (Optical Society of America, 2018). doi:10.1364/CLEO_AT.2018.JTh5C.2
29. Kang, K. *et al.* High-mobility three-atom-thick semiconducting films with wafer-scale homogeneity. *Nature* **520**, 656–660 (2015).
30. Braga, D., Gutiérrez Lezama, I., Berger, H. & Morpurgo, A. F. Quantitative Determination of the Band Gap of WS2 with Ambipolar Ionic Liquid-Gated Transistors. *Nano Lett.* **12**, 5218–5223 (2012).
31. Mlack, J. T. *et al.* Transfer of monolayer TMD $WS_2$ and Raman study of substrate effects. *Sci. Rep.* **7**, 43037 (2017).
32. Harris, N. C. *et al.* Efficient, compact and low loss thermo-optic phase shifter in silicon. *Opt. Express* **22**, 10487–10493 (2014).
33. Masood, A. *et al.* Fabrication and characterization of CMOS-compatible integrated tungsten heaters for thermo-optic tuning in silicon photonics devices. *Opt. Mater. Express* **4**, 1383–1388 (2014).
34. Chang, Y.-C., Roberts, S. P., Stern, B. & Lipson, M. Resonance-Free Light Recycling. (2017).
35. Muñoz, P. *et al.* Silicon Nitride Photonic Integration Platforms for Visible, Near-Infrared and Mid-Infrared Applications. *Sensors* **17**, (2017).
36. Timurdogan, E., Poulton, C. V., Byrd, M. J. & Watts, M. R. Electric field-induced second-order nonlinear optical effects in silicon waveguides. *Nat. Photonics* **11**, 200–206 (2017).
37. Wang, C., Zhang, M., Stern, B., Lipson, M. & Lončar, M. Nanophotonic lithium niobate electro-optic modulators. *Opt. Express* **26**, 1547–1555 (2018).
38. Kieninger, C. *et al.* Ultra-high electro-optic activity demonstrated in a silicon-organic hybrid modulator. *Optica* **5**, 739–748 (2018).
39. Gao, Y. *et al.* Large-area synthesis of high-quality and uniform monolayer $WS_2$ on reusable Au foils. *Nat. Commun.* **6**, 8569 (2015).

40. McCreary, K. M., Hanbicki, A. T., Jernigan, G. G., Culbertson, J. C. & Jonker, B. T. Synthesis of Large-Area $WS_2$ monolayers with Exceptional Photoluminescence. *Sci. Rep.* **6**, 19159 (2016).
41. Lee, J. *et al.* Crack-Release Transfer Method of Wafer-Scale Grown Graphene Onto Large-Area Substrates. *ACS Appl. Mater. Interfaces* **6**, 12588–12593 (2014).
42. Emani, N. K. *et al.* Electrical Modulation of Fano Resonance in Plasmonic Nanostructures Using Graphene. *Nano Lett.* **14**, 78–82 (2014).
43. Li, Y. *et al.* Measurement of the optical dielectric function of monolayer transition-metal dichalcogenides: $MoS_2$, $MoSe_2$, $WS_2$, and $WSe_2$. *Phys. Rev. B* **90**, 205422 (2014).

**Acknowledgements:** The authors thank Dr. Aseema Mohanty, Dr. Christopher T. Phare, Dr. Xingchen Ji, Dr. Steven A. Miller and Dr. Utsav Deepak Dave for fruitful discussions. The authors gratefully acknowledge support from Office of Naval Research (ONR) for award #N00014-16-1-2219, from Defense Advanced Research Projects Agency (DARPA) for award #HR001110720034 and #FA8650-16-7643. Research on tunable optical phenomena in TMD semiconductors was supported as part of Energy Frontier Research Center on Programmable Quantum Materials funded by the U.S. Department of Energy (DOE), Office of Science, Basic Energy Sciences (BES), under award #DE-SC0019443. C. P. and J.P. acknowledge funding from AFOSR (#FA9550-16-1-0031, #FA9550-16-1-0347). Y.Y. and L.C. acknowledges support from an EFRI award from NSF (#EFMA 1741693). S.H.C. was supported by the Postdoctoral Research Program of Sungkyunkwan University (2016). This work was done in part at the City University of New York Advanced Science Research Center NanoFabrication Facility, in part at the Cornell Nanoscale Facility, supported by the NSF award #EECS-1542081 and in part at the Columbia Nano Initiative (CNI) shared labs at Columbia University in the City of New York.

**Author contributions:** I.D. and M.L. conceived and proposed the TMD based photonic design and experiments. S.H.C, J.H. and D.N.B proposed the use of $WS_2$ for these photonic structures. C.P. and J.P. provided the MOCVD $WS_2$ film for the ionic liquid experiments, Y.Y. and L.C. provided the CVD $WS_2$ and $MoS_2$ film for the capacitive device experiments, and S.H.C and B.L. performed the TMD transferring and characterization (PL measurements). I.D. fabricated the composite photonic device, with assistance from S.H.C for TMD processing and development, G.R.B for the ITO development and M.A.T for the SU-8 cladding of the composite structures. I.D. performed and analyzed the optical measurements of the photonic devices. I.D. and G.R.B performed the capacitance measurement of these structures. I.D. and M.L. prepared the

## Methods

**Device Fabrication.**

**TMD preparation, transfer and patterning**- The PMMA/TMD stack was delaminated from the $SiO_2$/Si substrate by floating in a hot 1 M KOH solution with the PMMA side up. The PMMA/TMD stack is then rinsed in water a couple of times, before transferring it onto the SiN waveguides. After the transfer, the TMD clad waveguides are left to dry for a few hours before the PMMA was removed by soaking in acetone and rinsing in isopropanol. The transferred TMD is then patterned with HSQ/PMMA stack using electron beam lithography (EBL), followed by oxygen plasma and $CHF_3/O_2$ for etching residual PMMA and TMD. After the transfer, the TMD clad waveguides are left to dry for a few hours before the PMMA was removed by soaking in acetone and rinsing in isopropanol. The transferred TMD is then patterned with HSQ/PMMA stack using electron beam lithography (EBL), followed by oxygen plasma and $CHF_3/O_2$ for etching residual PMMA and TMD.

**Device Fabrication for SiN-TMD Ionic Liquid Photonic Devices.** We lithographically defined 1.3 μm wide waveguides on 330 nm high Silicon nitride (SiN), deposited using Low Pressure Chemical Vapor Deposition (LPCVD) at 800 °C and annealed at 1200 °C for 3 hours, on 4 μm thermally oxidized $SiO_2$, with a combination of e-beam lithography to define the waveguides and deep UV lithography to pattern the CMP marks. We etch the SiN waveguides and CMP patterns using $CHF_3/O_2$ chemistry in Oxford 100 Plasma ICP RIE, followed by the deposition of 600 nm of Plasma Enhanced Chemical Vapor Deposition (PECVD) silicon dioxide ($SiO_2$) on the waveguides. We planarize the $SiO_2$ to about 100 nm above the SiN waveguides using standard chemical planarization (CMP) techniques to create a planar surface for the transfer of the MOCVD $WS_2$ monolayer and to prevent the TMD film from breaking at the waveguide edges. The power splitter (combiner) at the input (output) of the MZI structure is designed using a $1 \times 2$ ($2 \times 1$) multimode interferometer (MMI). Due to the disproportionate ratio of the height (330 nm) to the width (1.3 μm) of the waveguide and sensitivity of the MMI structure, the MZI supports wavelength fringes from 1520 – 1630 nm for the TE mode and 1455 – 1480 nm for the TM mode.

Following the TMD preparation, transfer and patterning mentioned above, the metal contacts are lithographically patterned using the DUV mask aligner and 50/80 nm of Ti/Au is deposited using electron beam evaporation, followed by liftoff in acetone.

**Device Fabrication for SiN-TMD Capactitive Devices.** We lithographically defined 1 μm wide waveguides on 360 nm high Silicon nitride (SiN), deposited using Low Pressure Chemical Vapor Deposition (LPCVD) at 800 °C and annealed at 1200 °C for 3 hours, on 4 μm thermally oxidized $SiO_2$, using 248 nm deep ultraviolet lithography. We etch the SiN waveguides using $CHF_3/O_2$ chemistry in Oxford 100 Plasma ICP RIE, followed by the deposition of 600 nm of Plasma Enhanced Chemical Vapor Deposition (PECVD) silicon dioxide ($SiO_2$) on the waveguides. We planarize the $SiO_2$ to about 260 nm {40 nm} above the SiN waveguides using standard chemical planarization (CMP) techniques to create a planar surface for the transfer of the $WS_2$ {$MoS_2$} monolayer and to prevent the TMD film from breaking at the waveguide edges. The 260 {40 nm} nm $SiO_2$ layer prevents the lossy ITO layer of the capacitor from interacting with the optical mode too strongly. A 45 nm {5 nm} of thermal ALD alumina is deposited on top of $SiO_2$ to isolate the waveguides from the subsequent fabrication steps for the TMD-$HfO_2$-ITO capacitor. The TMD is prepared, transferred and patterned using the aforementioned process. The metal contacts to the TMD layer are then patterned using EBL and 30 nm Ti/ 50 nm Au is deposited using electron beam evaporation, followed by lift-off in acetone. 26 nm {37 nm} of thermal ALD Hafnia at 200 °C is then deposited to form the dielectric for the TMD-$HfO_2$-ITO capacitor. The other electrode of the capacitor is first patterned using EBL and then Indium Tin oxide (ITO) is sputtered at room temperature (24 °C), at a chamber base pressure of 12 mTorr, with an Argon flow of 30 sccm and 5 sccm oxygen flow, followed by lift-off in acetone. To reduce the resistivity of the sputtered ITO film, the substrate is heated to a temperature of 200°C and annealed in vacuum with the chamber pressure at $10^{-6}$ Torr, and an oxygen flow of 5 sccm for 30 minutes. We characterize the complex refractive index of ITO and thereby its doping using visible and infrared ellipsometry (Woollam VASE) and fit to a Drude-Lorentz relation (see Supplementary Section XI). Finally, the metal contacts to the ITO layer are patterned using EBL and 30nm Ti/ 50 nm Au is then deposited using e-beam evaporation, followed by lift-off in acetone. To define devices with SU-8 on the composite SiN-$WS_2$ waveguides, SU-8 is photolithographically patterned on the waveguide and developed, by spinning SU-8 to a thickness of 3.5 μm.

Note. The numbers in curly brackets {} indicate the dimensions for the composite SiN-MoS$_2$ devices.

**Experimental Setup.** We couple TE/TM polarized light from a tunable NIR laser (1450 nm – 1600 nm) to the SiN microring/MZI input using a tapered single mode fiber, which is then collected at the SiN bus/MZI output, using a similar tapered fiber. We record the microring/MZI transmission spectrum for different DC bias voltages applied across the WS$_2$-HfO$_2$-ITO capacitor, to measure the phase shift and absorption change. We normalize each of the MZI/ring transmission spectra first by the maximum output power across the measured wavelength range and then by the optical power spectral profile measured at the output of one of the MZI arms by coupling light to a similar SiN waveguide with only one MMI splitter at the input of MZI. The above normalization process eliminates the wavelength dependent optical loss experienced by the incident laser light in the tapered optical fiber, the fiber polarizer, SiN waveguide and the MMI, which are all designed to operate optimally at 1550 nm with the TE polarization. It also accounts for the wavelength dependent power fluctuations of the laser, if any.

For small-signal RF bandwidth measurement using the Vector Network Analyzer (VNA), we apply a DC bias voltage of 0 V across the WS$_2$-HfO$_2$-ITO capacitor, combined with an RF signal of -10 dBm with a source impendance of 50 Ω.

**Optical Sheet Conductivity of Monolayer TMD.** We use the 2D sheet conductivity model to extract the $\Delta n_{WS2}$ and $\Delta k_{WS2}$, as is commonly done when modeling graphene monolayers[42]. We extract the complex index change of the monolayer ($\Delta n + i\Delta k$) with carrier densities by comparing the measured $\Delta n_{eff}$ and $\Delta k_{eff}$ (Figure 1e)), with the simulated change obtained using COMSOL Multiphysics finite element model (FEM). We model the monolayer TMD as a conductive sheet with a surface charge density (J = $\sigma_s \cdot$ E) and complex optical conductivity[43] $\sigma_s = \sigma_R + j\sigma_I$. $\sigma_s(\omega)$ is related to the dielectric permittivity ($\varepsilon(\omega)$) through the equation

$$\sigma_s(\omega) = j\omega t_d \varepsilon_0 \left(\varepsilon(\omega) - 1\right) \tag{1}$$

where, $t_d$ defines the thickness of the monolayer TMD (0.85 nm ± 0.05 nm)[21]. We estimate the complex index change of the monolayer from the computed change in conductivity ($\Delta\sigma$) using the equation

$$\sigma_s(\omega) + \Delta\sigma = j\omega t_d \varepsilon_0 \left(\varepsilon(\omega) + \Delta\varepsilon - 1\right) \tag{2}$$

Where, $\Delta\varepsilon$ defines the change in dielectric permittivity ($\varepsilon(\omega)$), which is related to $\Delta n$ and $\Delta k$ through

$$\varepsilon(\omega) + \Delta\varepsilon = (n + \Delta n + i(k + \Delta k))^2 \tag{3}.$$

The change in the real part of effective index of the mode ($\Delta n_{eff}$) signifies a change in the imaginary part of the optical conductivity ($\Delta\sigma_I$), whereas the change in the imaginary part of the effective index ($\Delta k_{eff}$) of the mode is reflected in the real part of the optical conductivity ($\Delta\sigma_R$) through equation (1), (2) and (3). Since the change in complex effective index of the composite SiN-WS$_2$ waveguide is contingent on the spatial overlap between the propagating optical mode and the monolayer WS$_2$, the true measure of the efficiency of the phase modulation lies in the change of its in-plane optical sheet conductivity (S) and thereby its real and imaginary part of the refractive index ($\Delta n + i\Delta k$) with electrostatic gating. We calculate ($\Delta n_{WS2} + i\Delta k_{WS2}$) by modeling the monolayer as a boundary condition with surface current density ($J = \sigma_s \cdot E$), which eliminates the approximation of the thickness of the monolayer in our COMSOL model.

We quantify the mode overlap by replacing the surface current density ($J = \sigma \cdot E$) in the refractive index calculations, with a monolayer thickness ($t_d$) of 0.85 ± 0.05 nm thickness, and finding the surface integral to fraction of the energy flux ($E \times H$) of the mode in monolayer WS$_2$, compared to the total mode flux in the composite waveguide.

**Comparison of the electro-refractive change in WS$_2$ and MoS$_2$** - Even though the $\Delta n_{eff}$ for the air clad SiN-WS$_2$ composite waveguide (Figure 3c) is similar to the $\Delta n_{eff}$ of the air clad SiN-MoS$_2$ (Figure 4a), the extracted $\Delta n_{WS2}$ is higher than $\Delta n_{Mos2}$, due to the lower optical mode overlap of 0.016 % in the SiN-WS$_2$ waveguide, as compared to that of SiN-MoS$_2$ waveguide.